\documentclass[twocolumn,english,showpacs,superscriptaddress,prl]{revtex4}
\usepackage[T1]{fontenc}
\usepackage[latin9]{inputenc}
\usepackage{graphicx}
\usepackage{amssymb}
\usepackage{babel}

\begin{document}

\title{Magneto-electric photocurrent generated by direct inter-band
transitions in InGaAs/InAlAs two-dimensional electron gas}

\author{Junfeng Dai$^{\ast}$}
\affiliation{Department of Physics, The University of Hong Kong, Pokfulam Road, Hong
Kong, China}

\author{Hai-Zhou Lu$^{\ast}$}
\affiliation{Department of Physics, The University of Hong Kong, Pokfulam Road, Hong
Kong, China}
\affiliation{Centre of Theoretical and Computational Physics, The University of Hong
Kong, Pokfulam Road, Hong Kong, China}

\author{C. L. Yang}
\affiliation{Department of Physics, The University of Hong Kong, Pokfulam Road, Hong
Kong, China}
\affiliation{Department of Physics, Sun Yat-Sen University, Guangzhou, Guangdong, China}

\author{Shun-Qing Shen}
\affiliation{Department of Physics, The University of Hong Kong, Pokfulam Road, Hong
Kong, China}
\affiliation{Centre of Theoretical and Computational Physics, The University of Hong
Kong, Pokfulam Road, Hong Kong, China}

\author{Fu-Chun Zhang}
\affiliation{Department of Physics, The University of Hong Kong, Pokfulam Road, Hong
Kong, China}
\affiliation{Centre of Theoretical and Computational Physics, The University of Hong
Kong, Pokfulam Road, Hong Kong, China}

\author{Xiaodong Cui$^{\dag}$}
\affiliation{Department of Physics, The University of Hong Kong, Pokfulam Road, Hong
Kong, China}

\date{\today }

\begin{abstract}
We report observation of magneto-electric photocurrent generated via direct
inter-band transitions in an InGaAs/InAlAs two-dimensional electron gas
excited by a linearly polarized incident light. The electric current is
proportional to the in-plane magnetic field which unbalances the velocities
of the photoexcited carriers with opposite spins and consequently
generates electric current from a spin photocurrent. The observed light
polarization dependence of the electric current is explained microscopically
by taking into account of the anisotropy of the photoexcited carrier
density in wave vector space. The spin photocurrent can be extracted from
the measured current and the conversion coefficient of spin photocurrent
to electric current is estimated to be $10^{-3}$ $\sim $ $10^{-2}$ per Tesla.
\end{abstract}

\maketitle

Stimulated by the concept of non-magnetic semiconductor spintronics devices%
\cite{Awschalom-07np}, spin injection and detection by optical means have
attracted much attention\cite{Ganichev2002,Ganichev2003,Kato2004,
Wunderlich2005,Bhat2000,Stevens2003-prl,Hubner2003,Zhao2006-prl,Bhat2005,Zhao2005, Belkov2005,Ganichev2006,Li2006-apl,Cui2007,Zhou-07prb,photogavantic_Peking}%
. One way to generate spin current is to use linearly polarized optical
excitations in bulk III-V semiconductors and quantum wells (QW) with
asymmetric band structures induced by the strong spin-orbit coupling (SOC)
\cite%
{Bhat2005,Zhao2005,Belkov2005,Ganichev2006,Li2006-apl,Zhou-07prb,Cui2007}.
The left and right circular components of the linearly polarized light
generates the same amount of carriers with opposite spins and velocities,
leading to a spin photocurrent (SPC) accompanied with a null electric
current. The SPC may be generated by various microscopic optical absorption
mechanisms\cite{Tarasenko2005-JETP}. Because spin current carries neither
net charge nor magnetization, its measurement becomes a subtle issue\cite%
{Stevens2003-prl,Kato2004,
Wunderlich2005,Zhao2005,Zhao2006-prl,Valenzuela2006-nature,Cui2007,Appelbaum2007-prl,Wang2008-prl}.
Experimentally, the detection of the spin current has been realized by converting it to
either electric current or magnetization. As a linearly polarized light is
injected into a SOC material in the presence of an in-plane magnetic field,
the Zeeman splitting induces an imbalance of carriers of opposite spins in $\mathbf{k}$-space, resulting in electric photocurrent (EPC). The
field-induced EPC was systematically studied in the context of the
intra-band electron heating, during which abundant spin-dependent
excitation and relaxation processes are involved\cite%
{Ganichev2006,Belkov2005}.

\begin{figure}[tbph]
\centering \includegraphics[width=0.45\textwidth]{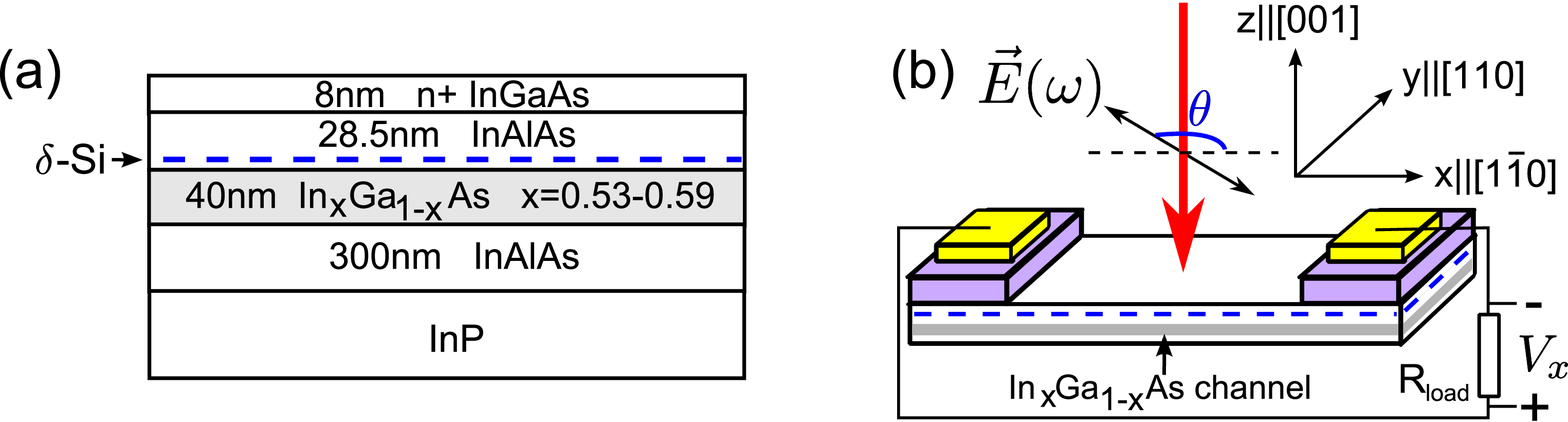} \centering
\includegraphics[width=0.47\textwidth]{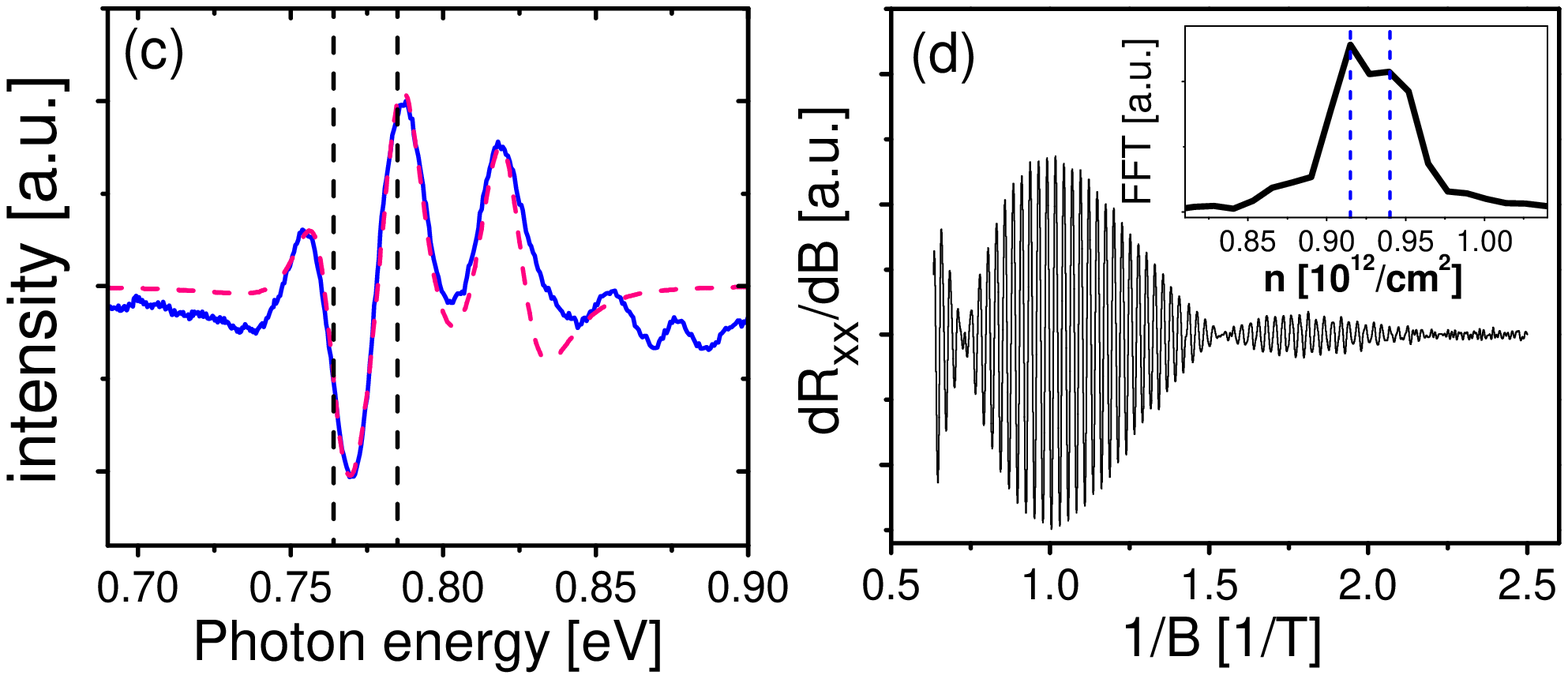}
\caption{(Color online) (a) [001]-oriented InGaAs/InAlAs quantum well. (b)
Illustration of the experiment setup. Linearly polarized light (downward
arrow) is normally incident with a polaziration angle $\theta$ with
respect to $x$-axis. (c) Photo-modulated differential reflectivity spectrum
(blue solid curve) at 77 K with the pump light of a DPSS laser at 532 nm.
Dashed curve is the spectrum analysis fitting curve\cite{reflectivity_analysis}. The two
lowest interband optical transitions below the photon energy (0.8 eV) are marked by the vertical dashed lines.
(d) Shubnikov-de Haas (SdH) oscillation
beating pattern of longitudinal magneto-resistance $R_{xx}$ at $T=2.5$ K.
Inset: FFT spectrum of $dR_{xx}/dB$ as a function of carrier concentration $
n $, indicating two spin-split subbands (vertical dashed lines).}
\label{fig:setup}
\end{figure}

In this Letter we report the observation of EPC generation via direct
interband transition by a linearly polarized light in the presence of an
in-plane magnetic field. The EPC is negligible in the absence of the field,
and is proportional to the in-plane magnetic field, which indicates the
conversion of spin current to electric current by unbalancing the
velocities of the photoexcited carriers with opposite spins. The Hall voltage in
the presence of a perpendicular magnetic field is used to estimate the
magnitude of the SPC when the light is incident on the edge of the sample.
The magneto-electric photocurrent and the Hall effect allow us to
self-consistently evaluate the magnitude of the SPC density from independent
approaches. The generation of the SPC and field-induced EPC can be well
elucidated by both the symmetry argument\cite{Belkov2005} and the
microscopic model with the anisotropic carrier density in $\mathbf{k}$-space
induced by interband optical excitation\cite{Bhat2005}.

The experiment is carried out on the modulation doped In$_{x}$Ga$_{1-x}$As/InAlAs QW grown along [001] direction, as shown in Fig. \ref{fig:setup}(a). The
40 nm InGaAs QW consists of graded indium composition from $x=0.53$ to $0.59$. The mobility and carrier (electron) density were measured as $\mu
_{m}=12000$ cm$^{2}$$\cdot $V$^{-1}$s$^{-1}$ and $n_{e}=2.2$ $\times $ $10^{12}$ cm$^{-2}$ at room temperatures by Hall measurements. The experiment
setup is shown in Fig. \ref{fig:setup}(b). A $1050$ $\mu$m long ($L$) and $50$ $\mu $m wide strip of 2DEG channel along [1\={1}0] and [110] directions
was fabricated by standard photolithography and wet etching.
Ni/Ge/Au/Ge/Ni/Au metal electrodes was deposited by electron-beam
evaporation and then annealed at 450$^{\circ }C$ for one minute to form
Ohmic contact. The linearly polarized light at 1mW normally sheds on the
channel through a $10\times $ objective lens and the light spot diameter is
approximately $D=10$ $\mu$m. The EPC $J_{x}$ passing through two terminals is
monitored via the voltage drop $V_{x}$ on a load resistor $R_{\mathrm{load}}=3.9$ k$\Omega $ with a lock-in amplifier.

The band gap of 0.764 eV is extracted from the photo-modulated differential
reflectivity spectrum\cite{reflectivity_analysis} performed at 77 K as shown
in Fig. \ref{fig:setup}(c). The pump source is a DFB packed laser diode at
1550 nm (0.80 eV) which excites interband transition only in the InGaAs
channel. Since the photon energy is well above the band gap and the two
lowest interband transition energies [marked by vertical dashed lines in Fig. \ref{fig:setup}(c)], the dominant optical absorption mechanism here is the direct
interband transition, i.e., an electron in a valance band is excited to a
conduction band. In the transition, the wave vector of electron $\mathbf{k}$
remains unchanged and its energy is increased by a photon energy $\hbar
\omega $. The intra-band transitions within the conduction bands,
in which higher-order scatterings by defects and phonons are needed to conserve
the momentum, are expected to have much smaller probabilities.

Zinc-blende heterostructures grown along [001] crystallographic direction,
like our device, usually have the $C_{2v}$ symmetry\cite{Belkov2005}. When
the measurement is performed along [1\={1}0] and [110] directions, the
dominant SOC is of the Rashba type. The SOC strength $\alpha $ can be
examined by the Shubnikov-de Hass (SdH) oscillation of the longitudinal
magneto-resistance $R_{xx}$ along the channel\cite{Nitta1997}. As shown in
Fig. \ref{fig:setup}(d), the SdH beating pattern of the derivative of $\mbox{R}_{xx}$ and the corresponding Fast Fourier transform (FFT) spectrum indicate that there exist two spin-split subbands due to the SOC. The Rashba
coefficient is estimated as $\alpha =4.3\times 10^{-12}$ eVm by $\alpha=\frac{\Delta n\hbar ^{2}}{m^{\ast }}\sqrt{\frac{\pi }{2(\overline{n}-\Delta
n)}}$, where $\overline{n}=0.93\times 10^{12}$/cm$^{2}$ and $\Delta
n=0.025\times 10^{12}$/cm$^{2}$ are the average value and the difference of
the two carrier concentrations extracted from the FFT spectrum\cite{SdH
oscillation}, and the effective mass $m^*=0.04m_e$ is assumed.

\begin{figure}[tbph]
\centering \includegraphics[width=0.47\textwidth]{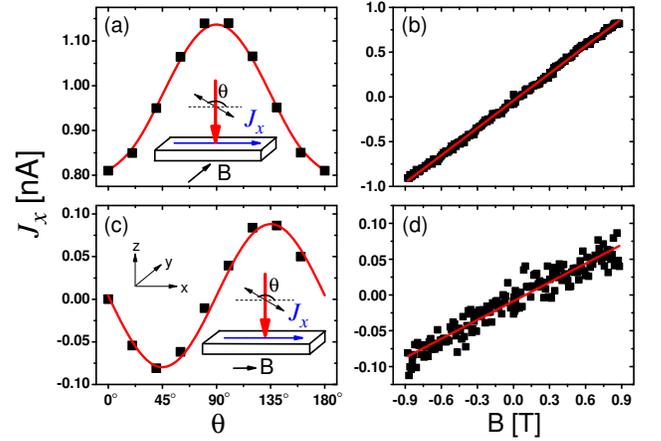}
\caption{(Color online) Electric photocurrent $J_{x}$ as a function of the
polarization angle $\theta$ and in-plane magnetic fields $B$, along $%
y$-axis [(a)-(b)], and along $x$-axis [(c)-(d)]. $B=0.9$ T in (a) and (c).
$\theta=20^{\circ}$ in (b), and $\theta=120^{\circ}$ in (d).}
\label{fig:Exp_I_B}
\end{figure}

In the absence of magnetic fields, the two states with opposite wave vectors
in each conduction band have opposite spins as required by the time-reversal
symmetry, resulting in a finite SPC but a null electric current\cite%
{Bhat2005,Tarasenko2005-JETP,Li2006-apl,Zhou-07prb}. As expected, we
measured a negligible EPC at zero bias without external magnetic field. To
explore the SPC, we apply an in-plane magnetic field to lift the spin
degeneracy by the Zeeman splitting. The SOC links the spin asymmetry to
the imbalance in $\mathbf{k}$-space and consequently an EPC arises. Fig. \ref%
{fig:Exp_I_B} shows the EPC $J_{x}$ induced by the 1550 nm light as a
function of in-plane magnetic fields along the $y$-direction [(a)-(b)] and
along the $x$-direction [(c)-(d)]. The EPCs are linearly proportional to the
magnetic field as shown in Figs. \ref{fig:Exp_I_B}(b) and \ref{fig:Exp_I_B}(d), and shows linear dependence on
the light intensity within the maximum laser power of 10 mW. The EPCs demonstrate
strong light polarization dependences as shown in Fig. \ref{fig:Exp_I_B} (a)
and (c). The relation between the EPC, the magnetic field, and the
polarization angle $\theta $ can be summarized by
\begin{equation}
J_{x}(B_{x},B_{y},\theta )=c_{0}B_{y}+c_{y}B_{y}\cos 2\theta +c_{x}B_{x}\sin
2\theta ,  \label{eq:JxB}
\end{equation}%
where $c_{0,x,y}$ are constants. According to Fig. \ref{fig:Exp_I_B}, $c_{0}$
overwhelms $c_{x,y}$, implying unpolarized light (also can be seen as
infinite linearly polarized lights) is enough for the observation.

Our observation is consistent with the symmetry argument of $C_{2v}$ group%
\cite{Belkov2005}. Because the $xz$ and $yz$ planes coincide with the mirror
reflection planes of the $C_{2v}$ group, the polar vectors (such as
velocity, current, electric field) along $x$ axis and the axial vectors
(such as spin and magnetic field) along $y$ axis transform according to the
representation B$_{1}$ of the $C_{2v}$ group, while the polar vectors along $%
y$ and the axial vectors along $x$ directions according to the
representation B$_{2}$\cite{Dresselhaus}. As a result, the EPC density in an
in-plane magnetic field can be phenomenologically written as\cite{Belkov2005}
$j_{\mu }=\sum_{\nu }(\chi ^{\mu \mu \nu \bar{\nu}}B_{\mu }E_{\nu }E_{%
\overline{\nu }}+\chi ^{\mu \bar{\mu}\nu \nu }B_{\bar{\mu}}E_{\nu }E_{\nu })$%
, where $\mu ,\nu $ run over $\{x,y\}$, and $\bar{\mu},\bar{\nu}=y$ if $\mu
,\nu =x$, and vice versa. $\chi $ is the fourth-rank pseudo tensor that
relates the EPC densities to the polarization electric field components $%
(E_{x},E_{y})\propto (\cos \theta ,\sin \theta )$ of the incident light and
magnetic fields $(B_{x},B_{y})$. From the above equation, one readily sees
the $\theta $-dependence of the EPC.

\begin{figure}[tbph]
\centering \includegraphics[width=0.4\textwidth]{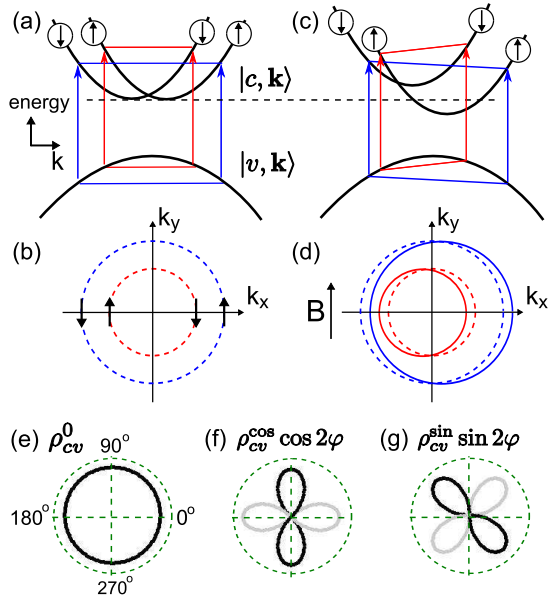}
\caption{(Color online) [(a) and (c)] Schematic process of direct optical
excitations from a valence band $|v,\mathbf{k}\rangle$ to the spin-split
conduction bands $|c=\pm,\mathbf{k}\rangle$ at magnetic field $\vec B$=0 (a)
and $\vec B$ along $y$-axis (c). [(b) and (d)] Constant energy contours
(CEC) of photoexcited conduction electrons, which are symmetric (dashed) at
$\vec{B}=0$, and shifted (solid) along $k_{y}$ ($k_{x}$) in the presence of
a magnetic field $B_{x}$ ($B_{y}$). The short arrows denote the spin orientations.
[(e)-(g)] Wave vector angle ($\varphi$)-dependence of photoexcited carrier density $\rho_{cv,\mathbf{k}}$ in Eq. (\ref{eq:rhock}) at $\vec{B}=0$. Only the
results on one CEC is shown. Dark (light) curves represent positive
(negative) values. $\rho_{cv}^{0}$ is always positive and overwhelms
$\rho_{cv}^{\cos/\sin}$ terms.}
\label{fig:schematic}
\end{figure}


Below we shall present a microscopic picture to further understand the
magnetic field and polarization dependences of the EPC in Eq. (1), by
considering the anisotropy of the photoexcited carrier density in $\mathbf{k}$-space\cite{Bhat2005}. We shall neglect the hole contribution from the
valence bands and focus on the conduction electrons, as the lifetime and
spin relaxation time of holes in the n-type are much shorter than those
of electrons. We shall illustrate our picture by considering two spin-split
conduction subbands $|c=\pm ,\mathbf{k}\rangle $ described by a Rashba
Hamiltonian with $C_{2v}$ symmetry\cite{note-C2v}, and a valence band $|v,%
\mathbf{k}\rangle $ [Fig. \ref{fig:schematic}(a)]. In $\mathbf{k}$-space,
the states involved in the interband transition are described by constant
energy contours (CEC), which are two perfect concentric circles split by the
SOC in the absence of magnetic fields [Fig. \ref{fig:schematic}(b)]. For
each state on the CEC, there exists another state with opposite velocity and
spin, therefore there is no EPC, but the difference between two CECs
caused by finite effective mass of holes and energy conservation leads to a
non-zero SPC. By applying the in-plane field, due to the SOC, the Zeeman
splitting of the spin-split conduction bands [Fig. \ref{fig:schematic}(c)]
shifts the CECs in $\mathbf{k}$-space along the direction perpendicular to
the B-field [Fig. \ref{fig:schematic}(d)]. This shift then breaks the
balance of velocities on the CECs and leads to a finite EPC. The EPC density
$j_{x}$ can be expressed as $j_{x}=-e\sum_{c,v}\int_{0}^{2\pi }d\varphi
\int_{0}^{\infty }kdk\delta (k-k_{cv})\rho _{cv,\mathbf{k}}v_{c\mathbf{k}%
}^{x}$, where $\varphi =\arctan k_{y}/k_{x}$, $-e$ is the electron
charge, $k_{cv}$ are the wave vectors on a CEC and $v_{c\mathbf{k}}^{x}$
denotes the $x$-axis velocity of $|c,\mathbf{k}\rangle $. To the linear
order in $\mathbf{B}$,
\begin{equation}
v_{\pm \mathbf{k}}^{x}=(\frac{\hbar }{m^{\ast }}k\pm \frac{\alpha }{\hbar }%
)\cos \varphi \mp \frac{\sin 2\varphi }{2\hbar k}h_{x}\mp \frac{\cos
^{2}\varphi }{\hbar k}h_{y},  \label{eq:vck}
\end{equation}%
where $h_{x(y)}$ is the Zeeman energy along $x(y)$ axis. The $\mathbf{k}$%
-dependent photoexcited carrier density $\rho _{cv,\mathbf{k}}$ may be
evaluated within the electric dipole approximation, and is found to be well
described by the following form in terms of the polarization angle $\theta $
at $\mathbf{B}=0$,
\begin{equation}
\rho _{cv,\mathbf{k}}=\rho _{cv}^{0}+\rho _{cv}^{\cos }\cos 2\varphi \cos
2\theta +\rho _{cv}^{\sin }\sin 2\varphi \sin 2\theta .  \label{eq:rhock}
\end{equation}%
The $\varphi $-dependence of $\rho _{cv,\mathbf{k}}$ is explicitly shown in
Fig. \ref{fig:schematic}[(e)-(g)]. The above expressions for $\rho _{cv,%
\mathbf{k}}$ and $v_{\pm \mathbf{k}}^{x}$ lead to Eq. (\ref{eq:JxB}) for $%
J_{x}$ \cite{note-microscopic}.

Since the field-induced EPC is well described by the microscopic picture, a
direct consequence is that a SPC is expected to exist in the absence of
magnetic field. The zero-field expectation value of spin velocity operator
along $x$ direction (defined as $\hat{\jmath}_{x}^{x/y}=\frac{1}{2}\{\hat{v}%
_{x},\hat{\sigma}_{x/y}\}$, where the superscripts denote the spin
orientation, and $\hbar/2$ is dropped for direct comparison with the current), can be found for states $|\pm ,\mathbf{k}\rangle $ as $%
(j_{x}^{x})_{\pm ,\mathbf{k}}=\pm \frac{\hbar }{2m^{\ast }}k\sin 2\varphi $,
$(j_{x}^{y})_{\pm ,\mathbf{k}}=(\mp \frac{\hbar }{m^{\ast }}k\cos
^{2}\varphi -\frac{\alpha }{\hbar })$. By comparing $(j_{x}^{x/y})_{\pm ,%
\mathbf{k}}$ with Eq. (\ref{eq:vck}), we can find that the SPC and the
field-induced EPC satisfy a rough but simple relation\cite{note-relation} $j_{\mu }^{\nu
}(B=0)/j_{\mu }(B\neq 0)\approx E_{\mathrm{K}}/h_{\mu }$,
where the Fermi kinetic energy $E_{\mathrm{K}}\approx 10^{-2}\sim 10^{-1}$
eV estimated from the carrier density $n_{e}$, and the Zeeman energy $h_{\mu
}$ is about $1.2\times 10^{-4}$ eV/Tesla (Land\'{e} g-factor $\approx $ 4
\cite{Nitta1997}). The EPC density $|j_{\mu }(B\neq 0)|$ is related to the
measured $J_{x}$ in Fig. \ref{fig:Exp_I_B} by $j_{\mu }(B\neq 0)=en_{e}\mu
_{m}J_{x}R_{\mathrm{load}}/L\approx 10^{-5}$ A/m, which assumes the
injection is balanced by the drift under the $V_{x}$-related built-in field.
Summarizing above gives a magnitude of the SPC density $|j_{\mu }^{\nu
}(B=0)|\approx 10^{-3}\sim 10^{-2}$ A/m for a 1 mW incident light, and a
conversion coefficient from SPC to field-induced EPC about $10^{-3}\sim
10^{-2}$ per Tesla.

\begin{figure}[tbph]
\centering \includegraphics[width=0.39\textwidth]{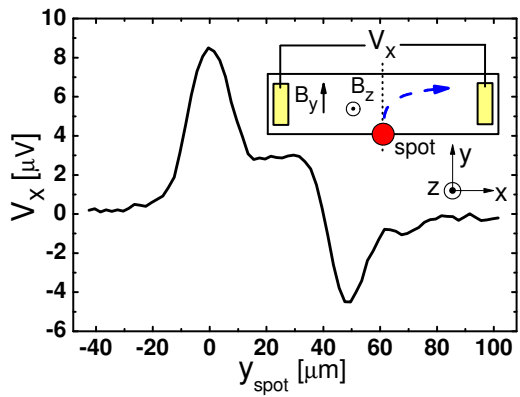}
\caption{(Color online) Hall voltage $V_{x}$ as a function of the spot
position $y_{\mathrm{spot}}$, in the presence of both in-plane and
out-of-plane magnetic fields ($B_{y}$ and $B_{z}$). The inset shows the
experimental setup, where a light is shed on the spot at edge of the sample.
The scanning path of $y_{\mathrm{spot}}$ is marked by the vertical dotted
line. The dashed arrow indicates the Hall current trajectory induced by $%
B_{z}$. $B_{y}$ = 0.86 T, $B_{z}$ = 0.15 T, the bar size is 50 $\times$ 1050
$\mu$m$^{2}$. }
\label{fig:hall}
\end{figure}

To further verify this estimate of the SPC density and conversion
coefficient, we employ a Hall measurement in which the light spot acts like
a current source in usual Hall bar setup. Besides an in-plane field $B_{y}$,
we also apply an out-of-plane magnetic field $B_{z}$, then scan the light
spot across the channel along $y$ axis. From the microscopic picture, it was
known that the light spot may generate photocurrents diffusing out of the
spot edge in all directions with the identical current density $j_{0}$. Note
that this isotropic $j_0$ yields no net electric current and is not the field-induced EPC. When the
spot is right focused on an edge of the sample [inset of Fig. \ref{fig:hall}%
], the anisotropic diffusion leads to a net current $Dj_{0}$ normal to the
edge and consequently produces a Hall voltage along $x$ direction under the
magnetic field $B_{z}$. Fig. \ref{fig:hall} shows $V_{x}$ as a function of
the spot position $y_{\mathrm{spot}}$ (the scanning path is marked by the
dotted line). The positive and negative peaks correspond to when the light spot is located at lower and upper edges of the channel,
respectively. If the spot is away from the edges, the diffusion of the
photocurrent is isotropic and consequently the Hall voltage vanishes. The
plateau in Fig. \ref{fig:hall} is attributed to the in-plane magnetic field
as those in Fig. \ref{fig:Exp_I_B}. This photoexcited Hall effect provides
another approach to estimate the SPC density. At the edges, $Dj_{0}$ is
related to $V_{x}$ by transverse Hall resistivity $\rho _{xy}\thickapprox
2B_{z}/\pi en_{e}$, then $j_{0}\approx \pi en_{e}V_{x}/2DB_{z}\approx
2.4\times 10^{-2}$ A/m at the peaks ($|V_{x}|\approx 6.5\mu $V) in Fig. \ref%
{fig:hall}. Approximately, a fraction of $\Delta n/\bar{n}\approx 0.027$
[Fig. 2(d)] of $j_{0}$ contributes to the SPC, giving rise to the estimated
SPC density $|j_{\mu }^{\nu }(B=0)|\approx j_{0}(\Delta n/\bar{n})\approx
0.7\times 10^{-3}$ A/m. Comparing it with the EPC signal (the plateau part)
owing to $B_{y}$, one could estimate the conversion coefficient of SPC to field-induced EPC at $\left\vert j_{\mu }(B\neq 0)/j_{\mu }^{\nu
}(B=0)\right\vert \backsimeq 1.7\times 10^{-2}$ per Tesla. These rough
estimates by the Hall effect agree with the low bound by the microscopic
model. Furthermore, $j_{0}$ can also be evaluated from the conversion of the
light power into the photocurrent. If the sample reflectance of 0.3,
photon-carrier yield of 30\% and the absorption coefficient of $9\times
10^{3}$cm$^{-1}$ are assumed, $|j_{\mu }^{\nu }(B=0)|\approx j_{0}(\Delta n/%
\bar{n})$ is around $10^{-2}$A/m and the conversion coefficient from SPC to field-induced EPC $\left\vert j_{\mu }(B\neq 0)/j_{\mu }^{\nu
}(B=0)\right\vert \backsimeq 10^{-3}$ per Tesla. These values are well
consistent with the high bound obtained by the above microscopic model.

In summary, the measurement of the magnetic-field-induced electric
photocurrent supports the signature of the spin photocurrent (SPC) induced
via the direct interband optical absorption in SOC materials. By two
independent approaches with the help of in-plane and out-of-plane magnetic
fields, the conversion coefficient of SPC to field-induced EPC is
estimated around $10^{-3}\sim 10^{-2}$ per Tesla. The experiment provides a
practical way to estimate the magnitude of the SPC density induced by
linearly polarized or unpolarized lights.

This work was supported by the Research Grant Council of Hong Kong under
Grant Nos. HKU 7013/08P , HKU 7041/07P, and HKU 10/CRF/08.

$^{\ast}$These authors contributed equally to this work.

$^{\dag}$ Email: xdcui@hku.hk


\begin{thebibliography}{99}
\bibitem{note-C2v} For simplicity, the microscopic picture in Fig. \ref%
{fig:schematic} assumes $C_{\infty v}$ symmetry, which can be generalized to
$C_{2v}$ by considering unequal Rashba coefficients along $x$ and $y$ axis.
For $C_{\infty v}$, $|c_{x}|=|c_{y}|$ in Eq. (\ref{eq:JxB}), which does not
hold for $C_{2v}$.
\bibitem{note-microscopic} Under the same principle of the $\varphi$
dependence, the linear order terms on $\mathbf{B}$ for $\rho_{cv,\mathbf{k}}$
also lead to almost the same amount of contribution to $J_{x}$ as those for $%
v_{c\mathbf{k}}$. For simplicity, this contribution is not explicitly
presented.
\bibitem{note-relation} The zero-field spin velocity $\propto \hbar
k/m^{\ast }$ while the magnetic field $h_{\mu }$ induced electron velocity $%
\propto h_{\mu }/\hbar k$, so their ratio roughly gives $\frac{\hbar
^{2}k^{2}}{2m^{\ast }}/h_{\mu }$, where the extra $2$ in front of $m^{\ast }$
is due to Ref. \cite{note-microscopic}. This relation is further interpreted
as \textquotedblleft kinetic energy over Zeeman energy".
\end{thebibliography}


\end{document}